\documentclass[reprint,aps,nofootinbib]{revtex4-1}
\usepackage{amsmath, amssymb}
\usepackage{tikz}
\usepackage[compat=1.1.0]{tikz-feynman}
\usepackage{siunitx}
\usepackage{array}
\usepackage{slashed}
\usepackage[hidelinks]{hyperref}

\newcommand{\jm}{J^\mu_{\mathrm{EM}}}
\newcommand{\curl}{\nabla \times}
\newcommand{\di}{\nabla \cdot}
\newcommand{\E}{\mathbf{E}}
\newcommand{\B}{\mathbf{B}}
\newcommand{\et}{\boldsymbol{\eta}}
\newcommand{\be}{\boldsymbol{\beta}}
\newcommand{\X}{\mathbf{X}}
\newcommand{\A}{\mathbf{A}}
\newcommand{\J}{\mathbf{J}}

\usepackage{xpatch}
\xpatchcmd\bibsection{19}{20}{}{}
\xpatchcmd\bibsection{\begingroup}{\vskip 20pt\begingroup}{}{}
\begin{document}

\title{Dark photon limits from magnetic fields and astrophysical plasmas}
\author{Giacomo Marocco \\ Rudolf Peierls Centre for Theoretical Physics, University of Oxford, Parks Road,\\ Oxford OX1 3PU, United Kingdom}
\begin{abstract}
    Dark photons may kinetically mix with our photon and modify Maxwell's equations. We examine their impact on the Lorentz force acting on the plasma in astrophysical magnetospheres. Solving the relevant magnetohydrodynamic equations, we show that any dark photon effect is negligible. The existing bound on dark photons with a mass below $10^{-14}$ eV, coming from \textit{Pioneer} 10 observations of Jupiter's magnetosphere, relies on such an effect and is thus unreliable. To rectify this, we use recent \textit{Juno} analyses to isolate any dark photon-induced modification of the Jovian magnetic field and hence place a numerically similar limit.   
\end{abstract}

\maketitle

\section{Introduction}

There may exist a light gauge boson called the dark photon (DP) $X_\mu$ that kinetically mixes with our photon $A_\mu$ \cite{holdomTwoChargeShifts1986}. The Standard Moded field content is uncharged under the corresponding gauge group, making such a boson difficult to detect as it mediates no long-range force, even if it is very light. 

The defining Lagrangian of the DP may be written as

\begin{equation}
\begin{split}
\mathcal{L} &\supset  -\frac{1}{4}\left(F_{\mu\nu}F^{\mu\nu} + X_{\mu\nu}X^{\mu\nu} \right) \\ &+ \frac{1}{2}m^2X^\mu X_\mu  + \jm A_\mu + \frac{\sin \alpha}{2} F^{\mu\nu} X_{\mu\nu},
\end{split}
\end{equation}
 where $X_{\mu \nu} = \partial_\mu X_\nu - \partial_\nu X_\mu$, $\sin \alpha$ parametrises the kinetic mixing, and $m$ is related to the DP mass. We remain agnostic as to the necessary UV completion of such a model, limiting ourselves to low energy scales where this approach is valid. To get to a basis with diagonal kinetic terms, we let $A_\mu \rightarrow \frac{ A_\mu}{ \cos \alpha}$ and $X_\mu \rightarrow X _\mu + \sin \alpha	 A_\mu $, so that the Lagrangian reads

\begin{equation}
\begin{split}
\mathcal{L} &\supset -\frac{1}{4}\left(F_{\mu\nu}F^{\mu\nu} + X_{\mu\nu}X^{\mu\nu} \right) \\ &+\tilde{e}J^\mu_{\mathrm{EM}}A_\mu + \frac{m^2}{2}\left( X^\mu X_\mu + 2 \chi X^\mu A_\mu + \chi^2 A^\mu A_\mu \right),
\label{eqn:lag}
\end{split}
\end{equation}
where $\chi \equiv \tan \alpha$ and $\tilde{e} \equiv e(1+\chi^2)^{1/2}$.  The Lagrangian still possesses the $U(1)$ gauge invariance of electromagnetism, despite appearances, in contrast to the Proca Lagrangian. The mass matrix is singular, containing a zero eigenvalue as dictated by gauge invariance; the other eigenvalue, the mass of the DP, is $m_X^2 = (1+\chi^2)m^2$. The model parameter space is then simply spanned by $\{m_X, \epsilon\}$.

There exist many constraints on the DP parameter space, see, e.g., \cite{jaeckelLowEnergyFrontier2010,Fabbrichesi:2020wbt} for a review of these. For ultralight DPs (below around \SI{e-10}{eV}), there are very strong constraints if the DP constitutes dark matter. If this is not the case, the bounds are weaker. A constraint comes from a Cavendish-type experiment \cite{PhysRevLett.26.721} exploring deviations from Coulomb's law \cite{bartlettLimitsElectromagnetic1988}. The  lack of spectral distortions measured by COBE \cite{Fixsen_1996} in the CMB places cosmological constraints  \cite{Caputo_2020,Garc_a_2020}. Observations of the magnetic fields around the Earth \cite{goldhaberTerrestrialExtraterrestrial1971a} also place limits \cite{bartlettLimitsElectromagnetic1988,Fischbach:1994ir, kloorLimitsNew1994}. Finally, Kloor et al. \cite{kloorLimitsNew1994} claim that measurements of Jupiter's magnetosphere, as reported in \cite{Smith305}, may bound dark photons at still lower masses.

We critically examine Kloor et al.'s recasting of the photon mass bound coming from Jupiter's magnetic field in terms of the DP model. We highlight the distinction between bounds that come from the observation of the forces on the plasma in a magnetosphere, and those that come from the the measured magnitude of the magnetic field ascribed to a dipole source.  The former, in the particular case of Jupiter, has been used to place a robust bound on the mass of the photon \cite{davisLimitPhoton1975}, but does not constrain DPs, as we will show. Furthermore, the latter is not enough to place constraints without including the external (non-intrinsic) contributions to the Jovian magnetic field, which formerly were not taken into account. 

The secure bound on the photon mass is derived from the non-observation of the effect of a massive photon on the plasma in the Jovian magnetosphere. As any photon mass increases, the size of the Lorentz force exerted by the magnetic field on a plasma increases, and at some point is in conflict with the observed flow.  We derive the DP-altered magnetohydrodynamic (MHD) equations and show that no such effect occurs and hence no recasting of the photon mass bound is possible. We thus turn to measurements from the \textit{Juno} mission \cite{2017SSRv} to obtain the size of any anomalous magnetic field that cannot be ascribed to the Jovian magnetosphere. This allows a reliable bound to be placed on very low mass DPs.

Incidentaly, we demonstrate that the most stringent bounds on the photon mass \cite{ryutovUsingPlasma2007} from observations of the approximate equilibrium of the solar wind around Pluto \cite{intermagnetic, doi:10.1063/1.1618536} do not constrain DPs either.

The layout of the paper is as follows. In section \ref{sec:Maxwell}, we present the modified form of Maxwell's equations, as well as the MHD equations in the quasistatic limit. In section \ref{sec:earthMagnet}, we present the bounds coming from careful observation of the Earth's magnetic field. In section \ref{sec:JupiterB}, we review the knowledge of Jupiter's magnetosphere that allows us to obtain an upper limit on the size of external magnetic fields around Jupiter, and thus to set a DP limit. In section \ref{sec:JupiterMHD}, we solve the relevant MHD equations to show that the forces on plasmas in the magnetosphere are largely unaffected by any DP, and hence their behaviour cannot be used to constrain DPs. We show in section \ref{sec:heliosphere} that the same conclusions are reached regarding the observation of currents in the far heliosphere around Pluto. Finally, we conlude in section \ref{sec:conclusion}. Appendix \ref{app:MDM} contains a derivation of the magnetic field created by a point-like MDM when Maxwell's equations are modified by the inclusion of a DP.

\section{DP-modifed Maxwell and MHD equations}
\label{sec:Maxwell}

In this section, we derive the equations governing the magnetic fields and currents within magnetospheres. In order to make contact with the common form of Maxwell's equations, we will work in non-relativistic notation. We write the regular electric and magnetic fields as $\E$ and $\B$, respectively, which may be written in terms of the scalar potential $\phi$ and vector potential $\A$ as 

\begin{subequations}
\begin{align}
\E &= - \nabla \phi - \partial_t \mathbf{A}, \\ 
\B &= \curl \mathbf{A}.
\end{align}
\end{subequations}
Similarly, we write the \textit{dark} electric and magnetic fields as $\et$ and $\be$, respectively. These are given in terms of their potentials as
\begin{subequations}
\begin{align}
\et &= - \nabla \psi - \partial_t \mathbf{X}, \\
\be &= \curl \mathbf{X}.
\end{align}
\end{subequations}

From the Lagrangian \eqref{eqn:lag}, we may derive the dark photon-modified Maxwell's equations
\begin{subequations}
\begin{align}
\di \E + \chi^2m^2\phi &= \tilde{e} \rho + \chi m^2 \psi, \\
\curl \B + \chi^2 m^2 \mathbf{A} &= \tilde{e} \mathbf{J} + \partial_t \E + \chi m^2 \mathbf{X}, \\
\curl \E &= -\partial_t \B, \\
\di \B &= 0.
\end{align}
\end{subequations}
Here, $\rho$ is the charge density and $\mathbf{J}$ is the charge current, both in units of $\tilde{e}$. Note that the homogeneous equations remain unchanged compared to the case with no dark photon.
The dark photon also satisfies its set of Dark Maxwell's equations
\begin{subequations}
\begin{align}
\di \et + m^2 \psi &= \chi m^2 \phi \\
\curl \be + m^2 \mathbf{X} &= \partial_t \et + \chi m^2 \mathbf{A} \\
\curl \et &= -\partial_t \be \\
\di \beta &= 0.
\end{align}
\end{subequations}
Additionally, the Lorentz force law is only modified by a renormalisation of the electric charge \cite{goldhaberTerrestrialExtraterrestrial1971a}, so we may write
\begin{equation}
\mathbf{F} = \tilde{e} (\E + \mathbf{v} \times \B),
\label{eqn:lorenzForce}
\end{equation}
where $\mathbf{v}$ is the velocity of a charged particle. Since the Lorentz force law \eqref{eqn:lorenzForce} is only modified by its overall scale, we may carry over all of the equations of kinetic plasma theory with the substitution $e \rightarrow \tilde{e}$.

To obtain a closed set of equations, we must further specify the equations of motion for the plasma. By the observation at the end of the previous paragraph, we get Ohm's law:
\begin{equation}
\tilde{e}\mathbf{J} = \sigma(\E + \mathbf{v} \times \B),
\end{equation} %
where $\sigma$ is the plasma conductivity. For plasmas in the solar system, we may take the large $\sigma$ limit; we further take the quasistatic limit to find the set of MHD equations that we will use:
\begin{subequations}
\begin{align}
(-\nabla^2+ \chi^2 m^2) \B &= \tilde{e} \curl \mathbf{J}  + \chi m^2 \be,\label{eqn:curlJ} \\
\curl \E &= -\partial_t \B,  \\
\E + \mathbf{v} \times \B &= 0, \\
(-\nabla^2 + m^2) \be &= \chi m^2 \B, \label{eqn:curlbe} \\
\curl \et &= - \partial_t \be, \\
\partial_t \rho + \di(\rho \mathbf{v}) &= 0, \\
\rho \frac{d \mathbf{v}}{t} &= - \nabla P + \tilde{e}\mathbf{J} \times \B.
\end{align}
\end{subequations}
Here $P$ is the fluid pressure, $\mathbf{v}$ the fluid velocity, $\mathbf{J}$ is still the current density in units of $\tilde{e}$, but we have \textit{switched notation} and $\rho$ is now the fluid density as the charge density does not appear any longer. 

\section{Earth's magnetic field}
\label{sec:earthMagnet}
The sources of magnetic fields around the Earth are well enough known that  bounds have been set on the photon's mass from measurements of the magnetic field caused by the magnetic dipole moment of the Earth \cite{Goldhaber:1968mt, Fischbach:1994ir}. These bounds are based on the technique of Schr\"odinger \cite{10.2307/20488455}. As the process for bounding dark photons will be similar, let us briefly recapitulate the logic of the massive photon case.

 If the photon has a mass $\mu$, the magnetic field caused by a dipole $\mathbf{D} = D \hat{z}$ is 
\begin{equation}
\B(\mathbf{x}) = \frac{De^{-\mu r}}{4\pi r^3}\Big[\left(1+\mu r + \frac{1}{3}(\mu r)^2\right)(3 \hat{z}\cdot \hat{\mathbf{x}}\hat{\mathbf{x}}-\hat{z})-\frac{2}{3}(\mu r)^2 \hat{z} \Big].
\label{eqn:massiveDipole}
\end{equation} At a constant radius, the terms in the first bracket of this equation simply effect a rescaling of the magnetic field compared to the massless case. The last term is a new effect though, and from the point of view of massless magnetostatics it appears like an external field antiparallel to the dipole direction. The ratio $S_\mu$ of the external field caused by the mass of the photon to the equatorial field strength is
\begin{equation}
S_\mu = \frac{2}{3}\frac{(\mu R)^2}{1+\mu R +(\mu R)^2/3},
\label{eqn:massRatio}
\end{equation}
where the measurement is taken at a radius $R$. There are a number of true external fields, which for the Earth are known. Subtracting these off leaves a bound on $S_\mu$ which translates into a bound on the photon mass $\mu \lesssim \SI{8 e-16}{eV}$ \cite{Fischbach:1994ir}. It is worth emphasising that the true external fields must be known to obtain a bound from this method. This is true of the Earth; it was not true of Jupiter at the time of early analyses that have been used in the literature, as stated in \cite{davisLimitPhoton1975}.  

 The case of a dark photon follows in the same way, as analysed in \cite{bartlettLimitsElectromagnetic1988,Fischbach:1994ir}
 (see Appendix \ref{app:MDM} for an alternate derivation of the field to all orders in $\chi$).
The resulting ratio $S$ of the new external field to the field strength at the equator is
\begin{equation}
S = \frac{2}{3}\frac{(\chi m_X R)^2  e^{-m_X R}(1+\chi^2)}{1+\chi^2e^{-m_X R}\left(1+m_X R + \frac{1}{3} (m_X R)^2 \right)}.
\label{eqn:ratioDP}
\end{equation}
The above ratio is bounded by $S \leq \SI{3.9e-4}{}$ at a distance $R = \SI{6.38e6}{m}= \SI{3.24e13}{eV^{-1}}$ \cite{Fischbach:1994ir}. The bounds that result are an order of magnitude more constraining than any other terrestrial experiment, and extend lower in mass than even those from COBE.

The limit that we reviewed here relies on the knowledge of the magnetic fields that arise from sources external to the Earth's magnetic dipole. Without this, it is impossible to ascribe any anomalous magnetic field to the presence of a dark photon, or equivalently a photon mass, rather than an unaccounted Standard Model process.

\begin{figure}[t!]
\centering
\includegraphics[width = 0.49\textwidth]{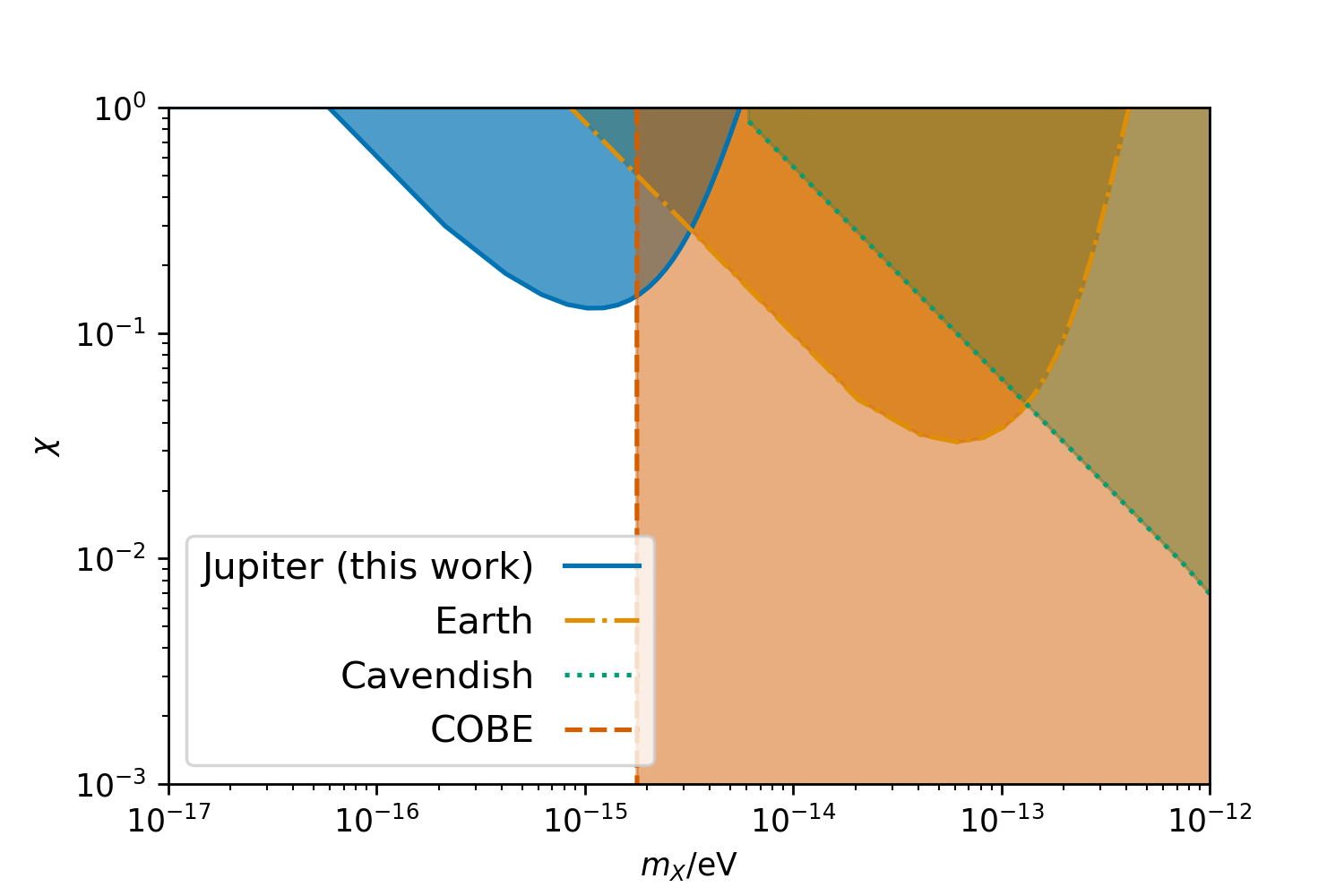}
\caption{Constraints on the dark photon parameter space. The existing constraints shown are: Earth, from the lack of anomalous external magnetic fields \cite{Fischbach:1994ir}; Cavendish, coming from the accuracy of Coulomb's law\cite{bartlettLimitsElectromagnetic1988}; COBE, from a lack of spectral distortions in the CMB \cite{Caputo_2020,Garc_a_2020}. The constraints we place from the size of any unaccounted external magnetic fields on Jupiter \cite{connerneyJovianMagnetodisc2020} are labelled as Jupiter (this work). }
\label{fig:bounds}
\end{figure}

\section{Jupiter's magnetic field}
\label{sec:JupiterB}
A number of missions have measured the magnetic field of Jupiter. \textit{Pioneer 10} took data in 1973 \cite{Smith1974ThePM}, and a year later the \textit{Pioneer 11} magnetometer \cite{1975SSI.....1..177A} obtained the most accurate measurement of the intrinsic, ``internal" Jovian dipole \cite{Acuna1976TheMM} until the \textit{Juno} mission \cite{https://doi.org/10.1002/2018GL077312}. The value obtained by the latter for the size of the dipole moment is 4.170$\,R_J\,$G, where $R_J = \SI{71492}{km}$ is the radius of Jupiter. The currents in the Jovian magnetosphere cause ``external" magnetic fields of order \SI{100}{nT}, and so at distances of order 10 $R_J$ the internal and external magnetic fields become comparable. Hence, in order to see any deviation from the predicted form of a MDM-sourced magnetic field at such radii, detailed knowledge of the external fields is required.

\textit{Voyager 1} measured not only the magnetic field around Jupiter \cite{nessMagneticField1979}, but critically also the currents and particles in its magnetosphere \cite{BRIDGE987,KRIMIGIS998, WARWICK995}. \textit{Voyager 2} obtained similar results for the field strength \cite{NESS966} and the magnetosphere's contents \cite{BRIDGE972,KRIMIGIS977, PEARCE991}. By fitting an empirical model \cite{connerneyModelingJovian1981} to the magnetic field observed by \textit{Juno}, the Jovian magnetosphere was shown to contain a large current sheet near the equatorial plane extending out from around 8 $R_J$ \cite{connerneyJovianMagnetodisc2020}. Using only \textit{Pioneer 10} data taken at radii of around 10 $R_J$, as done in \cite{kloorLimitsNew1994}, it is impossible to set a reliable bound, as one must account for the external distortion of the magnetic field produced by the internal Jovian dynamo.

Data from the \textit{Juno} magnetometer \cite{2017SSRv..213...39C} was analysed by Connerney et al. \cite{connerneyJovianMagnetodisc2020} to find the impact of the Jovian magnetodisc on the magnetic field. The resulting upper bound on the size of any unaccounted for external field, projected along the direction of the dipole and at a distance $R = 5 R_J = \SI{1.8e15}{eV^{-1}}$, is \SI{10}{nT}. In the absence of error bars in Connerney et al.'s analysis, we choose to double the size of this field to obtain a more conservative bound. The size of the intrinsic dipole at this radius is also needed, for which we use the analysis of \cite{https://doi.org/10.1002/2018GL077312} to obtain a value of \SI{3e3}{nT}. We thus bound the ratio \eqref{eqn:ratioDP} as $S \lesssim \SI{6e-3}{}$, as compared to \SI{e-1} used in \cite{kloorLimitsNew1994}. The resulting limits are shown in Fig. \ref{fig:bounds}.

One may worry that an empirical fit of the external magnetic field is not sufficient to extract a bound. To understand why, let us review the how this analysis work. Such a fit proceeds in two stages. First, the coefficients of the multipole moments of the internal field are measured by fitting to magnetic field close to Jupiter's surface, where external fields are highly subdominant \footnote{This analysis in fact shows that Jupiter is not well described as a dipole source, and the effect of higher order moments is significant, which also must be subtracted off.}. Then, using this internal model, the field at larger distances is fit to a particular magnetodisc model. In doing this fit, the effect of a DP may erroneously be included into the magnetodisc parameters if there is a degeneracy. If the time-variation of the magnetic field at fixed locations were available (as is the case on Earth, for example), then one could hope to isolate the time-independent effect of a DP from the external sources. However, since this is not the case, the external magnetic field ought to be determined from direct measurements of the plasma environment surrounding Jupiter. Models of the magnetosphere's charged particle distribution derived from the above fit in fact show consistency with the measured data by \textit{Voyager} 1 \& 2 to within a factor of two \cite{divineChargedParticle1983}, lending support to the conservative bound we have placed. A \textit{Juno} era consistency check would further support such a conclusion.

\section{MHD around Jupiter}
\label{sec:JupiterMHD}

The magnetic fields around Jupiter were analysed to obtain a bound on the photon mass $\mu \leq \SI{6e-16}{eV}$ \cite{davisLimitPhoton1975}, even without detailed knowledge of the surrounding currents. The procedure used to get this bound is different to that of the previous section, as emphasised in \cite{ryutovRoleFinite1997}. Let us briefly review how this was done.

For a massive photon, Amp\`ere's law is modified so that a dipolar magnetic field rescales the azimuthal current by a factor $1+(\mu r)^2$ compared to the case of a massless photon \cite{ryutovUsingPlasma2007}. Hence the force $\mathbf{F}=\mathbf{J}\times\B$ exerted on the plasma also increases by the same factor. Thus if the photon mass is too large, there can be no other balancing forces to achieve the observed equilibrium, and so such a scenario may be ruled out. This modified relation between currents and the magnetic field is the source of the above limit.

How does a dark photon modify the currents in Jupiter's magnetic field? We follow Ryutov \cite{ryutovUsingPlasma2007} and use a purely kinematic model for the standard electromagnetic interactions. We consider a situation in which the magnetic field on some surface of radius $r_0$ is known to be a dipole, so that we may write 
\begin{equation}
\B = B_0 \cos \theta \hat{r}.
\end{equation}
on this surface. Then, as we move away, the vanishing divergence of $\B$ implies 
\begin{equation}
\B = B_0 r_0^2 \frac{\cos \theta}{\mathbf{r}^2}\hat{r}.
\end{equation}
We now would like to solve equation \eqref{eqn:curlbe} to find the dark magnetic field in the presence of such a source. The particular solution to this equation may be written in Cartesian co-ordinates as 
\begin{equation}
\be(\mathbf{x}) = \frac{\chi m^2}{4\pi}\int d^3 \mathbf{\tilde{x}}\frac{e^{-m|\mathbf{\tilde{x}}|}}{|\mathbf{\tilde{x}}|}\B(\mathbf{x} + \mathbf{\tilde{x}}).
\label{eqn:partSol}
\end{equation}
We will solve the integral perturbatively in two different limits. As in the case of a massive photon, we ignore the homogeneous solution as unlike in regular magnetostatics there is no spatially homogeneous solution, and we are just left with unphysical exponential solutions.
\subsection{Large mass limit}
\label{sec:largeMassJup}
We first proceed in the limit $|\mathbf{x}|m \gg1$.
The strategy is to now expand $\B(\mathbf{x} + \mathbf{\tilde{x}})$ around the point $\mathbf{x}$ and then evaluate the integral \eqref{eqn:partSol}. The exponential damping factor will tame the divergent pieces in the Taylor series of $\B$ as long as we are in the specified limit. The result is
\begin{equation}
\be(\mathbf{x}) = \chi B_0r_0^2\left( \be(\mathbf{x})^{(0)} + \frac{\be(\mathbf{x})^{(2)}}{(mr)^2} + \frac{\be(\mathbf{x})^{(4)}}{(mr)^4}+ \ldots\right), \label{eqn:beLargem}
\end{equation}
where 
\begin{equation}
\begin{split}
\be(\mathbf{x})^{(0)} &=  \frac{\cos\theta}{r^2}\hat{r}, \\
\be(\mathbf{x})^{(2)}  &= -2 \left(\frac{\cos\theta}{r^2} \hat{r} +\frac{\sin\theta}{r^2} \hat{\theta} \right), \\
\be(\mathbf{x})^{(4)} &= -8 \left(\frac{\cos\theta}{r^2} \hat{r} + 2 \frac{\sin\theta}{r^2} \hat{\theta} \right). 
\end{split}
\end{equation}

In order to find how this dark magnetic field affects the charge current, we plug the result \eqref{eqn:beLargem} into \eqref{eqn:curlJ}. Note that the zeroth order piece in \eqref{eqn:beLargem} cancels what looks like a photon mass term in \eqref{eqn:curlJ}. This means that the only effect of a dark magnetic field is higher order in $mr$ than one would expect from a naive comparison between the Proca Lagrangian and the Lagrangian we consider. This may be seen explicitly in the solution for $\mathbf{J}$ which is
\begin{equation}
\mathbf{J} = B_0 r_0^2 \frac{\sin\theta}{r^2}\left( 1+\chi^2 + \frac{4\chi^2}{(mr)^2} + \ldots\right).
\end{equation}
If the change in the current induced by the dark magnetic field were to have a noticeable effect, we must have at least $\chi>1$, since this result only holds if $mr\gg 1$. However, $\chi>1$ is outside of the realm of applicability of the tree-level model of dark photon electromagnetism that we consider, and so no consistent constraints can be obtained.

\subsection{Small mass limit}

\label{sec:smallMassJup}
Let us now work in the other limit where the characteristic scale of observation $R$ is much less than the Compton wavelength of the dark photon $m R \ll 1$. The solution to \eqref{eqn:curlbe} may then be approximated as 
\begin{equation}
\be \simeq \chi (mR)^2\B,
\end{equation}
and so to leading order in $mR$ \eqref{eqn:curlJ} is just
\begin{equation}
(-\nabla^2+ \chi^2 m^2) \B = \tilde{e} \curl \mathbf{J} + \ldots,
\end{equation}
which is of the same form as that arising from the Proca Lagrangian with a photon mass $\mu = \chi m$. We may thus recycle the known results to show that the force exerted on the plasma by the current is
\begin{equation}
\tilde{e}\mathbf{J}\times \B = \frac{B_0^2 r_0^2 \sin\theta\cos\theta}{4\pi r^5}\left(1+(\chi m R)^2 + \ldots\right).
\end{equation}
This would be incompatible with observations if the extra term in brackets were somewhat larger than one; in the limit $mR \ll 1$, this requires a $\chi$ larger than one, which is again outside of the scope of classical physics. No bound can be placed on dark photons from current forces around Jupiter in this limit of the dark photon mass.

We have now shown that the observation of a steady state in the neighbourhood of Jupiter places no consistent constraint on the dark photon coupling, in contrast to the case of a massive photon. 

\section{MHD in the far heliosphere}
\label{sec:heliosphere}

The strongest reliable constraint on the photon mass $\mu \lesssim \SI{10e-18}{eV}$ comes from consideration of the solar wind in the far heliosphere \cite{ryutovUsingPlasma2007}. Can one derive a constraint on dark photons in a similar manner? 

The starting point is the model of solar winds first put forward by Parker \cite{parkerDynamicsInterplanetary1958}. The components of the magnetic field when we no longer neglect the azimuthal component of the magnetic field that comes from the Sun's rotation, as is necessary at large distances $r \gg r_0$, are given by
\begin{equation}
\begin{split}
B_r &= B_0r_0^2 \frac{\cos\theta}{r^2},\\
B_\theta &= 0, \\
B_\phi &= B_0r_0^2 \frac{\omega}{v_m}(r-r_0)\frac{\sin\theta}{r^2},
\end{split}
\end{equation}
where $\omega$ is the angular velocity of the sun and $v_m$ is the outward velocity of the solar wind. In the far heliosphere, the magnetic field is almost purely azimuthal. We now follow the same procedure as section \ref{sec:JupiterMHD} to see what effect there is on the current.

\subsection{Large mass limit}
Carrying out the same expansion as in section \ref{sec:largeMassJup}, we write 
\begin{equation}
\be(\mathbf{x}) =  \be(\mathbf{x})^{(0)} + \frac{\be(\mathbf{x})^{(2)}}{(mr)^2} + \ldots,
\end{equation}
and solve \eqref{eqn:curlbe} perturbatively to find
\begin{equation}
\begin{split}
 \be(\mathbf{x})^{(0)} & = \chi \B, \\
 \be(\mathbf{x})^{(2)} &= - 2 \chi \B.
\end{split}
\end{equation}
Thus \eqref{eqn:curlJ} gives
\begin{equation}
\tilde{e} \curl \J = \frac{2\B}{r^2}(1+\chi^2 + \ldots),
\end{equation}
which implies that $\J$ has $r$- and $\theta$-components of rough size 
\begin{equation}
J \sim \frac{B_\phi}{r}(1+\chi^2)^{1/2},
\end{equation}
with an associated force causing equatorial compression and radial deceleration of magnitude
\begin{equation}
F \sim \frac{B_\phi^2}{r}(1+\chi^2)^{1/2}.
\end{equation}
If this force were bigger than the force $\rho v^2/r$, then the velocity would not be radial and constant, as is observed. We thus need $F<\sigma \rho v^2/r$ where $\sigma$ is some fudge factor to account for our crudeness; this is the same limit used by \cite{ryutovUsingPlasma2007} to derive a limit on the photon mass. Translating this into dark photon parameters, we find the constraint
\begin{equation}
1+ \chi^2 \lesssim 10^{3/2};
\end{equation}
the bound is trivially satisfied for any perturbative coupling.

\subsection{Small mass limit}

As in section \ref{sec:smallMassJup}, to leading order in the small mass limit, we again have
\begin{equation}
(-\nabla^2+ \chi^2 m^2) \B = \tilde{e} \curl \mathbf{J} + \ldots.
\end{equation}
This is solved by 
\begin{equation}
\tilde{e}\J = 2\frac{B_\phi}{r}\left(1+(\chi m r)^2\right)\hat{r},
\end{equation}
and again we see the dark photon correction is always negligible.

\section{Conclusions}
\label{sec:conclusion}

The magnetic field produced by a MDM receives noticeable corrections when the photon kinematically mixes with a massive DP. Using this fact and knowledge about the true external fields, we placed constraints on dark photons whose Compton wavelength is of the order of the size of the Jupiter. These bounds extend further down in DP mass than any other (apart from those that make assumptions about the nature of dark matter). Specifically, we have $\chi \gtrsim 10^{-1}$ for $m_X \sim \SI{e-15}{eV}$. This made use of a fit of the magnetic field to an empirical model of the Jovian magnetodisc. This bound could be made more compelling by checking for good consistency between this model and the direct measurements of the surrounding plasma with forthcoming \textit{Juno} data. A full comparison of the functional form of the magnetic field around Jupiter with that predicted by Maxwell theory -- taking into account the effect of the surrounding plasma -- would likely give stronger bounds, but is beyond the scope of this work.

We have shown that bounds on the photon mass coming from the observation of current flow patterns in astrophysical magnetospheres do not constrain DPs, in contrast to massive photons. For large DP masses, where one might hope to have some effect, the leading order terms cancel once one consistently solves the relevant MHD equations. DPs with small masses fare no better, causing no significant alteration of the force felt by the plasma.

\section*{Acknowledgements}
I thank John Wheater for insightful conversations and suggestions, and Subir Sarkar and John Wheater for comments on a draft. I am grateful to Cressida for supporting me during this work.
\appendix

\section{The magnetic field from a magnetic dipole moment}

\label{app:MDM}
We would like to calculate the magnetic field arising from a point magnetic dipole moment, in analogy with \eqref{eqn:massiveDipole}. The relevant equations we need to solve are 
\begin{subequations}
\label{eqn:MHDearth}
\begin{align}
(- \nabla^2 + \chi^2 m^2)\A  &= \tilde{e} \J + \chi m^2 \X ,\\
(-\nabla^2 + m^2)\X &= \chi m^2 \A,
\end{align}
\end{subequations}
where we work here in Coulomb gauge $\di \A = 0$ and note that in the quasistatic limit this also implies $\di \mathbf{X} = 0$. Upon Fourier transforming \eqref{eqn:MHDearth}, these differential equations become algebraic and we may solve for the vector potential to find
\begin{equation}
\mathbf{A}(x) = \tilde{e} \int \frac{d^3 \mathbf{k}}{(2\pi)^3}\frac{\mathbf{\tilde{J}}(\mathbf{k})e^{i \mathbf{k}\cdot\mathbf{x}}}{k^2+\chi^2m^2\frac{k^2}{k^2+m^2}},
\label{eqn:vectorPotFT}
\end{equation}
where $\tilde{\J}(\mathbf{k})$ is the Fourier transform of $\tilde{e}\J(\mathbf{x})$. Approximating the magnetic dipole as coming form a point-like source at the origin, we have $\J(\mathbf{x})= - \mathbf{D} \times \nabla \delta(\mathbf{x})$, where $\mathbf{D} = D \hat{z}$ is the magnetic dipole moment. We then have $\tilde{\J}=-i\mathbf{D}\times \mathbf{k}$.

We now need to evaluate integrals of the form 

\begin{equation}
\mathbf{I} =  \int \frac{d^3 \mathbf{k}}{(2\pi)^3}\frac{\mathbf{k}e^{i \mathbf{k}\cdot\mathbf{x}}}{k^2+\chi^2m^2\frac{k^2}{k^2+m^2}},
\end{equation}
which is most readily done in spherical co-ordinates. If we rotate our axes so that the polar axis of the momentum integral aligns with $\hat{\mathbf{x}}$, we see that the $x-$ and $y$-components have vanishing azimuthal angle $\phi$ integrals. The polar integral needed is found to be
\begin{subequations}
\begin{align}
\int_0^\pi d \theta \,e^{i k r \cos\theta} \cos \theta \sin\theta &= 2i\frac{-kr \cos(kr)+\sin(kr)}{(kr)^2}.
\end{align}
\end{subequations}
After integrating $\phi$ from 0 to $2\pi$, we may extend the remaining even integral over the whole real axis and evaluate it by the Residue Theorem as
\begin{widetext}
\begin{equation}
\frac{i}{r^2}\int^\infty_{-\infty}\frac{dk}{(2\pi)^2}\,k\frac{-kr \cos(kr)+\sin(kr)}{k^2+\chi^2m^2\frac{k^2}{k^2+m^2}} = \frac{i}{4\pi r^2}\left( \frac{1+\chi^2e^{-mr\sqrt{1+\chi^2}}}{1+\chi^2}+ \frac{mr\chi^2 e^{-mr\sqrt{1+\chi^2}}}{\sqrt{1+\chi^2}} \right).
\end{equation}
\end{widetext}
If we now rotate back to the original co-ordinate system, we find that $\mathbf{I}$ points in the same direction as $\mathbf{x}$.
Recalling \eqref{eqn:vectorPotFT}, which was $\A = - i \mathbf{D} \times \mathbf{I}$, we find 
\begin{equation}
\A(\mathbf{x})= \frac{D}{4\pi r^2}\left( \frac{1+\chi^2e^{-mr\sqrt{1+\chi^2}}}{1+\chi^2}+ \frac{mr\chi^2 e^{-mr\sqrt{1+\chi^2}}}{\sqrt{1+\chi^2}} \right) \hat{\phi},
\end{equation}
for $\mathbf{D}$ in the $z$-direction. Taking the curl of this equation gives us the magnetic field sourced by a magnetic dipole moment when a dark photon coupling is present:
\begin{equation}
\begin{split}
\B(\mathbf{x}) = &\frac{D}{4\pi r^3(1+\chi^2)^2 } \\ \cdot \Big\{  &\Big[1+\chi^2e^{-m_X r}\left(1 +m_X r + \frac{1}{3} (m_X r)^2 \right)\Big] \Big( 3 \mathbf{\hat{x}}\cdot \hat{z} \mathbf{\hat{x}}- \hat{z} \Big)  \\
&-\frac{2}{3}(\chi m_X r)^2  e^{-m_X r}(1+\chi^2)	  \hat{z}\Big\}.
\end{split}
\end{equation}

\bibliography{references}

\end{document}